\documentclass[final,5p,times,twocolumn]{elsarticle}

\usepackage{graphicx}
\usepackage{epstopdf}
\usepackage{dcolumn}
\usepackage{bm}
\usepackage[english]{babel}

\usepackage[colorlinks=true,linkcolor=black, citecolor=blue, urlcolor=blue]{hyperref}

\usepackage{amsmath,amssymb,bm,bbm}
\usepackage{graphicx}
\usepackage{multirow}
\usepackage{color}
\usepackage[normalem]{ulem}
\usepackage{tikz}

\usepackage{float}
\restylefloat{figure}

\usepackage{mathrsfs}

\newcommand{\ket}[1]{ | #1 \rangle }

\newcommand{\overlap}[2]{\langle #1 | #2 \rangle}
\newcommand{\elmx}[3]{\langle #1 | #2 | #3 \rangle}

\begin{document}

\begin{frontmatter}



  \title{Exact solutions of the nuclear shell-model secular problem: Discrete Non-Orthogonal Shell Model within a Variation After Projection approach}


\author[first]{Duy-Duc Dao}
\author[first]{Frédéric Nowacki}
\affiliation[first]{organization={Université de Strasbourg, CNRS, IPHC UMR7178},
            addressline={23 rue du Loess, F-67000 Strasbourg}, 
            country={France}}

\begin{abstract}
  We investigate the capacity of non-orthogonal many-body expansions in the resolution of the nuclear shell-model secular problem. Exact shell-model solutions are obtained within the variational principle using non-orthogonal Slater determinants as the variational ansatz. These results numerically prove the realization of the Broeckhove-Deumens theorem on the existence of a discrete set of non-orthogonal wavefunctions that exactly span the full shell-model space for low-lying states of interest. With the angular-momentum variation after projection, pairing correlations are shown to be fully captured by Slater determinants as exemplified in the backbending phenomenon occurred in $^{48}$Cr. The resulting discrete non-orthogonal shell model developed in such variation after projection method is further examined in the case of $^{78}$Ni, an exotic doubly magic nucleus at the edge of currently feasible diagonalization limits. Its ground state binding energy is shown to converge to a lower value than the largest large-scale shell-model diagonalization ever done by the conventional tridiagonal Lanczos method, revealing an outstanding performance of non-orthogonal Slater determinantal wavefunctions to describe the eigensolutions of shell-model Hamiltonians. \\

\end{abstract}




\end{frontmatter}




\section{Introduction}
\label{introduction}

In quantum mechanics, one can in principle know all the physical properties of a many-body interacting system by solving the associated eigenvalue problem of the respective Hamiltonian. Even though in practice one eventually works with a finite-dimensional Hilbert space, in general this is always a highly non-trivial question because the many-body space size increases quickly with the number of degrees of freedom. A key for making this task computationally manageable is to find relevant dynamical quantities (e.g. the wave function, many-body densities or Green functions) and their representations.

For the description of strongly correlated fermionic systems such as atomic nuclei, the seminal works of Hill, Wheeler, Griffin~\cite{Hill1953_PhysRev.89.1102,Griffin1957_PhysRev.108.311} and Peierls, Yoccoz~\cite{Peierls1957,Yoccoz1957A_Moments_Inertia,PEIERLS1962} opened an appealing strategy using continuous non-orthogonal representations of the wave function, whose properties and practical realization were extensively studied in~\cite{Pizza1977_PhysRevC.15.1477,Pizza1980_PhysRevC.21.425,Deumens1979,Deumens1981}. In particular, the Broeckhove-Deumens theorem, pointed out in~\cite{Deumens1979}, states that, for a separable Hilbert space $\mathscr L^2$ and a closed subspace $\mathscr H = \overline{\mathrm{span} \: \mathit\Gamma} \subseteq \mathscr L^2$ where $\mathit\Gamma$ is a continuous, dense set of non-orthogonal states, there exists a discrete countable non-orthogonal subset $\mathit\Gamma_0 \subset \mathit\Gamma$ such that $\mathscr H = \overline{\mathrm{span} \: \mathit\Gamma_0}$. The theorem cannot tell us about the nature of the sets $\mathit\Gamma$, $\mathit\Gamma_0$ nor the way to find them, it however provides the formal justification for a discrete representation of the wave function $\psi = c_1\phi_1 + \dots + c_k\phi_k + \dots$ employing a non-orthogonal basis $\{\phi_k\}$. In other words, the theorem guarantees the existence of such a non-orthogonal basis whose determination thus depends on the specific many-fermion system under consideration.

In the case of nuclear systems, the nuclear Shell Model (SM) constitutes an ideal playing ground for exploring the capacity of non-orthogonal wave functions in realistic calculations of nuclei. This is especially advantageous as the direct full-fledged diagonalization of effective shell-model Hamiltonians in physically sound and well-defined valence spaces is nowadays feasible in many places of the nuclear chart thanks to enormous algorithmic advances and computational powers~\cite{Caurier2005,BigStickJOHNSON20132761,NuShellXBROWN2014,KShellSHIMIZU2019372,NOWACKI2021}. Indeed, the combination between non-orthogonal many-body expansions of the wavefunction and the symmetry restoration concept~\cite{Sheikh2021} has been the core of various approximation methods which were developed within the shell-model context~\cite{Schmid1984A_PhysRevC.29.291,Schmid1984B_PhysRevC.29.308,Schmid2004,Otsuka1996_PhysRevLett.77.3315,OTSUKA2001MCSM,Horoi2009_PhysRevC.80.034325,Horoi2009_PhysRevC.79.014311,MCSM2012,Gao2015_PhysRevC.92.064310,Gao2017_PhysRevC.95.064307,Gao2018_PhysRevC.98.021301,QVSM2021_PhysRevC.103.014312,Bally2019SD_PhysRevC.104.054306,Bally2019PF_PhysRevC.100.044308,DNO2022_PhysRevC.105.054314}. A central question in using non-orthogonal expansions is whether it is possible to achieve the exact shell-model diagonalization solution, for example, of the ground state in which the absolute binding energy and correlations were found to be difficult to be fully captured~\cite{Horoi2009_PhysRevC.80.034325,MCSM2012,Gao2018_PhysRevC.98.021301,Bally2019PF_PhysRevC.100.044308,DNO2022_PhysRevC.105.054314}. This issue was addressed within the angular-momentum projection after variation (PAV)~\cite{Otsuka1996_PhysRevLett.77.3315,OTSUKA2001MCSM,Horoi2009_PhysRevC.80.034325,Horoi2009_PhysRevC.79.014311,Bally2019SD_PhysRevC.104.054306,Bally2019PF_PhysRevC.100.044308,DNO2022_PhysRevC.105.054314} and also in the angular-momentum variation after projection  (VAP)~\cite{Schmid2004,MCSM2012,Gao2018_PhysRevC.98.021301,QVSM2021_PhysRevC.103.014312} approaches, and is still an open question. Without the consideration of symmetry restoration, similar questions of the energy convergence are also the concern for earlier and recent works either within a shell-model space~\cite{Faessler1969,HoKim1970_PhysRevLett.25.123,Hokim1970Erratum_PhysRevLett.25.782.3,HoKim1971_PhysRevC.4.1077} or in energy density functional frameworks~\cite{Robin2016_PhysRevC.93.024302,Robin2017_PhysRevC.95.044315,Matsumoto2023_PhysRevC.108.L051302}. In these latter many-body methods, it is also not clear whether an exact solution of shell-model type can be achieved. To put the question in a more general term in the sense of the Broeckhove-Deumens theorem, can we find a discrete non-orthogonal subset $\mathit\Gamma_0$ that exactly spans the full shell-model space $\mathscr H_{SM}$ for low-lying states of interest ?

In this work we investigate these issues and obtain exact shell-model solutions using symmetry-projected non-orthogonal Slater determinantal ansatz in a variety of physical regimes, hence numerically prove the realization of the Broeckhove-Deumens theorem. With the incorporation of spin-variation after projection, we study the capacity of Slater determinants in the description of nuclear pairing correlations. We then examine their performance in the binding energy convergence of $^{78}$Ni, an exotic nucleus representing current state-of-the-art shell-model diagonalizations~\cite{Nowacki2016_PhysRevLett.117.272501,Taniuchi2019_Ni78Nature}.

\section{Discrete Non-Orthogonal Shell Model}
\label{method}
As pointed out in~\cite{Deumens1979}, the basis selection technique suggested by Etienne Caurier (to be described later) allows to variationally determine a discrete non-orthogonal family $\it\Gamma_0$ out of a predefined continuous set $\it\Gamma$. The Broeckhove-Deumens theorem elucidates this technique on a firm theoretical ground, which inspired us to define the "Discrete Non-Orthogonal Shell Model" (DNO-SM)~\cite{DNO2022_PhysRevC.105.054314}. The latter work presents an implementation of the continuous set $\it\Gamma$ based essentially on the triaxial quadrupole deformation $(\beta,\gamma)$ following the classic generator coordinate method~\cite{Hill1953_PhysRev.89.1102,Griffin1957_PhysRev.108.311,Peierls1957,Yoccoz1957A_Moments_Inertia} as applied within the context of atomic nuclear systems~\cite{RingSchuck1980}. This construction of $\it\Gamma$, nonetheless, does not imply \textit{a priori} the completeness stipulated by the Broeckhove-Deumens theorem (cf. related discussions in~\ref{APP}). 

In order to pursue this question of completeness, we present here two extensions of the DNO-SM. The first one within the projection after variation approach includes both deformed configurations and their particle-hole excitations. The second one is developed in the angular-momentum variation after projection~\cite{Schmid1984A_PhysRevC.29.291,MCSM2012,Gao2018_PhysRevC.98.021301} using non-orthogonal Slater determinants as the variational wave function ansatz. The underlying Hamiltonian is composed of a one-body part $t_{ij}$ and a two-body effective interaction $V_{ijkl}=-V_{jikl}=-V_{ijlk} = V_{jilk}$
\begin{equation}
  \hat H = \sum_{ij} t_{ij} c^\dagger_i c_j + 
  \frac 1 4 \sum_{ijkl} V_{ijkl} c^\dagger_i c^\dagger_j c_l c_k
\end{equation}
where $c^\dagger_i,c_i$ are the creation and annihilation operators satisfying the standard anticommutation rules and the indices $i,j,k,l$ label the spherical harmonic oscillator (SHO) single-particle orbitals in the valence space.

Within the PAV approach, we expand the nuclear wave function $\ket{\psi^{\pi JM}_n}$ (labeled by $n$) of good angular momentum $J$ and parity $\pi$ as the superposition of the non-orthogonal deformed Hartree-Fock (HF) Slater determinants $\ket{\phi_q}$ and complemented further by their associated many-particle many-hole ($N$p$N$h) excitations $\ket{\phi_q(N\mathrm{p}N\mathrm{h})}$ which were not considered in our previous study~\cite{DNO2022_PhysRevC.105.054314}
\begin{equation}
\begin{aligned}
  \ket{\psi^{\pi JM}_n} = \sum_{q,K}C^{\pi J}_{n,qK} \mathcal P^J_{MK}\:P^\pi\ket{\phi_q} 
    + \sum_{q,K,N}C^{\pi J}_{n,qK,N} \mathcal P^J_{MK}\:P^\pi
  \ket{\phi_q(N\mathrm{p}N\mathrm{h})}.
\end{aligned}
\label{Psi_NpNh}
\end{equation}
The intrinsic HF states $\ket{\phi_q}$, labeled by a chosen generator coordinate generically denoted by $q$, and their many-particle many-hole excitations $\ket{\phi_q(NpNh)}$ are both projected onto good quantum numbers $J/\pi$ by the angular momentum/parity projection operators $\mathcal P^J_{MK}/P^\pi$~\cite{RingSchuck1980}. $M,K$ are the angular momentum projections onto the laboratory and intrinsic frames respectively. The mixing amplitudes $C^{\pi J}_{n,qK},C^{\pi J}_{n,qK,N}$ form the eigenvectors of the generalized eigenvalue equation
\begin{equation}
  \mathcal H^{\pi J} \: C^{\pi J}_n = E^{\pi J}_n \: \mathcal N^{\pi J} \: C^{\pi J}_n,
  \label{secular_problem}
\end{equation}
with $O^{\pi J}_{qK,q'K'} = \elmx{\phi_q}{\hat O \mathcal P^{J}_{KK'}P^\pi}{\phi_{q'}}\:(\hat O = \hat H,\mathbbm 1)$ being the Hamiltonian ($\mathcal H^{\pi J}$) and norm ($\mathcal N^{\pi J}$) matrices and $E^{\pi J}_n$ the energy level.

The determination of deformed Hartree-Fock states $\ket{\phi_q}$ is proceeded in the same manner as in our previous work~\cite{DNO2022_PhysRevC.105.054314} with the Caurier basis selection technique: by setting up a generator coordinate mesh points $\{q_k,k=1,2,3...\}$, we take the Hartree-Fock minimum $\ket{\phi_{q_1}}$ as the first state. The second state $\ket{\phi_{q_2}}$ is chosen such that the energy level $E^{\pi J}_n$ obtained from the Hill-Wheeler equation~\eqref{secular_problem} in the 2-dimensional space spanned by $\{\ket{\phi_{q_1}},\ket{\phi_{q_2}}\}$
be a minimum. And we continue similarly for the third state $\ket{\phi_{q_3}}$ until the convergence. In this way we generate a minimal set of $q$-constrained Hartree-Fock states. Starting from this minimal set, we repeat the same process to find the necessary $N$p$N$h excitations until the convergence. This procedure thus constructs the discrete non-orthogonal set $\mathit\Gamma_0^{(PAV)}$ for a targeted state $J^\pi_n$, which we will refer to as DNO-SM(PAV).

The VAP approach is more complicated~\cite{Schmid2004,RingSchuck1980} since it aims to build up the intrinsic states by a direct minimization of the projected energy functional without referring to generator coordinates. By the Thouless theorem~\cite{RingSchuck1980,THOULESS1960225}, we characterize the non-orthogonal Slater determinantal states as
\begin{equation}
  \begin{aligned}
    \ket{\phi_q} = \mathcal N_0 e^{\sum_{ij}Z^{(q)}_{ij}a^\dagger_ia_j}\ket{\phi^{(q)}_0}, \: q=1,2,3,\dots,
  \end{aligned}
  \label{Thouless}
\end{equation}
which is specified, up to a normalization constant $\mathcal N_0$, by a skew-symmetric complex matrix $Z^{(q)}$ and a randomly predefined reference Slater state $\ket{\phi^{(q)}_0}$. The symmetry-conserving nuclear state of good angular momentum $J$ and parity $\pi$ is now given by
\begin{equation}
\ket{\psi^{\pi JM}_n} = \sum_{q,K}C^{\pi J}_{n,qK} \mathcal P^J_{MK}\:P^\pi
\ket{\phi_q}.
\label{Psi}
\end{equation}
Using the trial variational ansatz~\eqref{Psi}, we can now minimize the projected energy functional under the normalization condition
\begin{equation}
  E^{\pi J}_n =  
  \frac{\elmx{\psi^{\pi JM}_n}{\hat H}{\psi^{\pi JM}_n}}{\overlap{\psi^{\pi JM}_n}{\psi^{\pi JM}_n}}, \:
  \overlap{\psi^{\pi JM}_n}{\psi^{\pi JM}_n} = 1.
\label{Evap}
\end{equation}
The first variation with respect to the mixing amplitudes $C^{\pi J}_{n,qK}$ leads to the secular equation~\eqref{secular_problem}. The second variation with respect to $\ket{\phi_q}$, using the Thouless parametrization~\eqref{Thouless}, leads to the generalized Brillouin condition
\begin{equation}
  \begin{aligned} 
    \frac{\displaystyle\sum_{qK,q'K'} C^{*\pi J}_{n,qK}C^{\pi J}_{n,q'K'} \elmx{\phi_q}{a^\dagger_j a_i\big(\hat H - E^{\pi J}_n\big)\mathcal P^J_{KK'}P^\pi}{\phi_{q'}}}{\displaystyle\sum_{qK,q'K'}C^{*\pi J}_{n,qK}C^{\pi J}_{n,q'K'} \elmx{\phi_q}{\mathcal P^J_{KK'}P^\pi}{\phi_{q'}}} = 0, \: \forall q,i,j
  \end{aligned}
  \label{Second_variation}
\end{equation}
where $a^\dagger_\lambda = \sum_i D^{(\lambda)}_i c^\dagger_i$ and $D^{(\lambda)}_i$ the single-particle state matrix for the Slater state $\ket{\phi_{q}}$. The condition~\eqref{Second_variation} means the nuclear state $\ket{\psi^{\pi J}_n}$ is stationary with respect to 1p1h configuration mixings of all Slater states $\{\ket{\phi_q},q=1,2,...\}$. This corresponds to the resonating Hartree-Fock approach originally proposed by Fukutome~\cite{Fukutome1988} in quantum chemistry. In nuclear physics, the VAMPIR~\cite{Schmid2004} and subsequent frameworks~\cite{MCSM2012,QVSM2021_PhysRevC.103.014312} use the few determinant approach which is a numerically less demanding variant by optimizing one state at a time, i.e. varying the last added $q+1$ state while keeping unchanged the previous $1,2,...,q$ variationally optimized ones. The Brillouin condition~\eqref{Second_variation} in that case will not invoke a summation over $q$ states. 

In our present work, to optimize the $q$-Slater configuration, we combine these two extreme cases in the following way: we keep unchanged the previous $1,2,...,q_0$ variationally optimized states and perform the resonating Hartree-Fock approach for the last added $q\geq 1$ ones. In practice we use a quasi-Newton method~\cite{Neff2008,Liu1989,Nocedal2006} to carry out the minimization of the projected energy~\eqref{Evap}. The non-orthogonal set $\mathit\Gamma_0^{(VAP)}=\{\ket{\phi_q},q=1,2,3,\dots\}$ is thus dynamically generated from the Hamiltonian itself in such a way that parity and rotational symmetries are automatically incorporated for low-lying states $J^\pi$ of interest. We will refer to this hybrid approach as DNO-SM(VAP).

\section{Analysis of dynamical correlations}
\label{sect3}
With the universal sd (USDB) interaction~\cite{Brown2006_PhysRevC.74.034315}, light nuclei are a realistic testing ground for variational methods~\cite{Otsuka1996_PhysRevLett.77.3315,Horoi2009_PhysRevC.79.014311,Gao2018_PhysRevC.98.021301,Bally2019SD_PhysRevC.104.054306,DNO2022_PhysRevC.105.054314}. Closely resembling to a SU(3) rotor picture of Elliot~\cite{NilssonSU32015_PhysRevC.92.024320}, typical open-shell N=Z nuclei as $^{20}$Ne, $^{24}$Mg or $^{28}$Si are well approximated by the configuration mixing of several deformed HF states using the quadrupole coordinates $(\beta,\gamma)$~\cite{DNO2022_PhysRevC.105.054314}. There is, however, an amount of ground state binding energy of order $\sim 100-600$ keV not fully captured in previous  studies~\cite{Horoi2009_PhysRevC.80.034325,Gao2018_PhysRevC.98.021301,Bally2019SD_PhysRevC.104.054306,DNO2022_PhysRevC.105.054314}, which we want to investigate here in relation with the Broeckhove-Deumens theorem. Note that this mismatch also manifests in heavier systems~\cite{Horoi2009_PhysRevC.80.034325,Horoi2009_PhysRevC.79.014311,Bally2019PF_PhysRevC.100.044308}, pointing to dynamical correlations beyond the static quadrupole deformation~\cite{Bender2008_EPJAS}. 

\begin{figure}[H]
  \centering
  \includegraphics[scale=1.6]{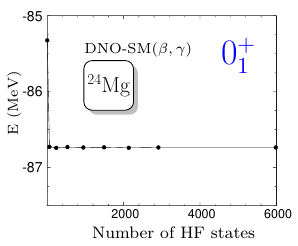}
  \caption{Ground state binding energy convergence with the number of $(\beta,\gamma)$-constrained Hartree-Fock (HF) states. $\beta$ is the deformation parameter and $\gamma$ measures the triaxiality~\cite{RingSchuck1980}.\label{convergence}}
\end{figure}

Table~\ref{Exactsd} presents the comparison of the ground state energy between the $(\beta,\gamma)$ configuration mixing without and with $N$p$N$h excitations (DNO-SM(PAV)), the variation after projection calculations (DNO-SM(VAP)) and the exact shell-model (SM) solutions. In the 2nd column, using the triaxial $(\beta,\gamma)$ configurations, the largest energy difference compared to the SM solution is in typical cases of $^{24}$Mg, $^{28}$Si. The number of HF states, found by the Caurier basis selection technique, are respectively $16$ and $28$. However, this is not a problem of small or large basis dimension. It is rather a question about the nature of correlations. To illustrate this point, Fig.~\ref{convergence} shows the energy convergence of $^{24}$Mg with the number of $(\beta,\gamma)$-constrained HF states.  The converged value (with $\sim 6000$ HF states in the sextant $0^\circ \leq \gamma \leq 60^\circ$) is found to be the same as the one obtained with $16$ HF states. This result, higher than the exact SM energy, thus implies i) the triaxial quadrupole deformation coordinates generate strictly a subspace $\mathcal E_{def}$ of the full shell-model space $\mathscr H_{SM}$ (cf.~\ref{APP}), and ii) the missing correlations are of a different nature. In the 3rd column of Table~\ref{Exactsd}, the incorporation of $N$p$N$h excitations form the complement of $\mathcal E_{def}$, bringing the extra energy to match the exact solution. Interestingly, the necessary excitations are found, by Caurier's basis selection procedure, to be associated with $N=2,4,6,8$ and composed of time-reversed conjugate states coupled to $K=0$, reminiscent of the formation of a pair condensate structure in Talmi model~\cite{TALMI1971_SeniorityModel} but built on top of deformed time-reversal invariant HF states. This feature suggests a manifestation of pairing correlations that play in the ground state wave function. Their impact on, e.g. E2 transition observables, remains albeit very perturbative due to a strong dominance of the quadrupole SU(3) regime.
\begin{table}[H]
\centering
\scalebox{0.85}{
        \begin{tabular}{ccccc}
          \hline\hline 
          & \multicolumn{2}{c}{DNO-SM(PAV)} & \multirow{2}{*}{\textcolor{black}{DNO-SM(VAP)}} & \multirow{2}{*}{Exact SM} \\ \cline{2-3}
          & $(\beta,\gamma)$ & $(\beta,\gamma)$+\textcolor{blue}{$N$p$N$h} & \\ 
          \hline
          \multirow{2}{*}{$^{20}$Ne} & $-40.35736$  & \textcolor{blue}{$-40.47233$}  &  \textcolor{red}{$-40.47231$}     &\textcolor{violet}{\textbf{$-40.47233$}} \\
                                    & $7$ & $51$ & $3$   &  $640$ \\
          \multirow{2}{*}{$^{24}$Mg} & $-86.73278$  & \textcolor{blue}{$-87.10428$}  &  \textcolor{red}{$-87.10405$}     &\textcolor{violet}{\textbf{$-87.10445$}} \\
                                    & $16$ & $975$ & $16$   &  $28503$ \\
          \multirow{2}{*}{$^{28}$Si} & $-135.21742$ & \textcolor{blue}{$-135.85891$} &  \textcolor{red}{$-135.86003$}     &\textcolor{violet}{\textbf{$-135.86073$}} \\
                                    & $28$ & $4255$ & $45$   &  $93710$ \\
          \multirow{2}{*}{$^{26}$Al} & $-$ & $-$ &  \textcolor{red}{$-105.74901$}     &\textcolor{violet}{\textbf{$-105.74934$}} \\
                                    & $-$ & $-$ & $24$   &  $26914$ \\
          \hline\hline
      \end{tabular}}
      \caption{Comparison of the ground state binding energy (in MeV) between the DNO-SM(PAV) without (2nd column) and with $N$p$N$h excitations (3rd column), DNO-SM(VAP) (4th column) and exact Shell Model (Exact SM) calculations for typical nuclei in the sd shell using the USDB interaction~\cite{Brown2006_PhysRevC.74.034315}. The first line indicates the binding energy in MeV and the second line shows the number of basis states in each calculation.\label{Exactsd}}
\end{table}
\noindent Remarkably, the DNO-SM(VAP) approach (4th column) captures all correlations to recover exactly the SM solution at the cost of a very few states. The energy matching in the case of the odd-odd nucleus $^{26}$Al is also excellent, indicating that the spin-variation after projection with Slater determinants is capable to account for correlations in even- and odd-systems equally well. These results thus constitute an ample numerical evidence of the Broeckhove-Deumens theorem on the completeness of non-orthogonal Slater determinants for shell-model valence spaces. They also show that the spin-variation after projection is strictly equivalent to performing $N$p$N$h excitations upon spherical or deformed reference states. In other words, it represents an efficient way to exactly incorporate correlations brought by these excitations in the full space of huge dimensions.

\section{Description of pairing correlations with non-orthogonal Slater determinants}
\label{sect4}
The preceding examples have shown that for even-even systems the ground state energy difference between the exact SM solution and the configuration mixing of $(\beta,\gamma)$-constrained HF states has its origin in the formation of a pair condensate of time-reversed partners coupled to $K=0$, a signature of pairing correlations. With the Kuo-Brown (KB3) effective interaction in the pf shell, we examine now the relevance of non-orthogonal Slater determinants in a textbook shell-model description of the rotational band in $^{48}$Cr~\cite{Caurier2005,POVES1998203} where the effect of pairing correlations is more spectacular.
\begin{figure}[H]
  \centering
  \includegraphics[scale=0.43]{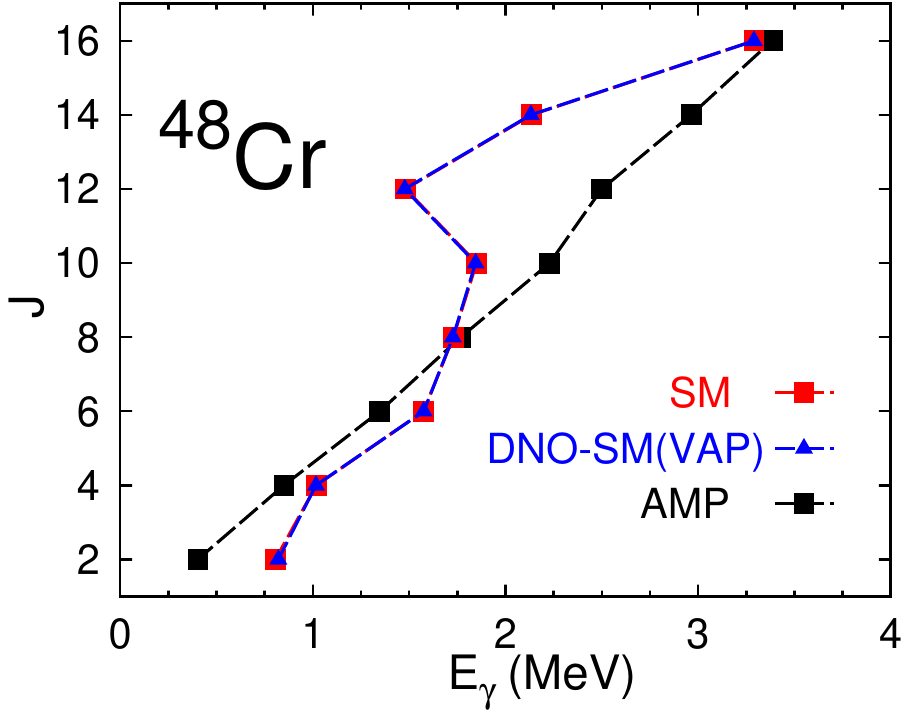}
  \caption{$\gamma$-excitation energies ($E_\gamma$) versus angular momentum ($J$) comparison between the Angular Momentum Projection (AMP) of the HF minimum, the DNO-SM(VAP) and the exact SM calculations. \label{Cr48backbend}}
\end{figure}
Fig.~\ref{Cr48backbend} shows the comparison of the DNO-SM(VAP) and exact SM calculations in the description of the backbending occurrence in $^{48}$Cr. The calculations are performed with the minimization of the $q$-Slater configurations simultaneously for each state of the \textit{yrast} band. The number of Slater determinants in the set $\it\Gamma_0^{(VAP)}$ is given in Table~\ref{VAPCr48} together with the comparison of the absolute energies. As seen in Fig.~\ref{Cr48backbend}, the two curves becomes indistinguishable with a perfect matching. This indicates a remarkable performance of non-orthogonal Slater determinants to capture pairing correlations. It is worth noting that, as pointed out in previous studies~\cite{Caurier2005,POVES1998203}, the proton-neutron pairing is found to be mandatory if one uses the Hartree-Fock-Bogoliubov treatment. Here our results show that it is possible to fully incorporate proton-neutron pairing correlations using non-orthogonal Slater determinants while conserving the particle numbers. This example shows again a firm numerical proof of the Broeckhove-Deumens theorem.
\begin{table}[H]
\centering
\scalebox{0.9}{      
      \begin{tabular}{cccc}
          \hline\hline
          \multirow{2}{*}{$J^\pi_n$} & \multirow{2}{*}{$\Gamma_0^{(VAP)}$} & \multicolumn{2}{c}{$E$ (MeV)}  \\\cline{3-4}
                        &                     & DNO-SM(VAP) & SM \\
          \hline
          $0^+_1$ & $50$ &  $-32.953$  & $-32.953$ \\
          $2^+_1$ &  $57$   &  $-32.135$     & $-32.148$ \\
          $4^+_1$ &  $56$   &  $-31.117$     & $-31.130$ \\
          $6^+_1$ &  $48$   &  $-29.541$     & $-29.555$ \\
          $8^+_1$ &  $43$   &  $-27.813$     & $-27.825$ \\
          $10^+_1$ & $30$   &  $-25.968$     & $-25.976$ \\
          $12^+_1$ & $28$   &   $-24.491$    & $-24.495$ \\
          $14^+_1$ & $19$ &  $-22.359$     & $-22.359$ \\
          $16^+_1$ & $12$ &  $-19.068$     & $-19.068$ \\
          \hline\hline
      \end{tabular}      
      }      
      \caption{Realization of the Broeckhove-Deumens Theorem in $^{48}$Cr. Comparison of the absolute energy ($E$) for the \textit{yrast} band. \label{VAPCr48}}
\end{table}
In a wider context, as noted by A. Bohr and Ben R. Mottelson~\cite{Mottelson1975_VolII}, the substantial deviations from the rotational law $J(J+1)$ characterize a strong influence of pair correlations, resulting in a significant modification of the rotational pattern. In light systems, $^{48}$Cr is thus such a typical case under the strong regime of pairing dominance, in complete contrast to situations encountered in well-deformed heavier-mass nuclei. For instance, in the transfermium region, the experimental data reveals a quasi-perfect agreement with the rotor spectrum $J(J+1)$ in many cases~\cite{Reiter1999_PhysRevLett.82.509,Greenlees2012_PhysRevLett.109.012501,Seweryniak2023_PhysRevC.107.L061302}. This feature signifies a much weakened pairing effects in heavy well-deformed nuclei compared to light ones. In view of these characteristics, the use of Slater determinants under a variation after projection framework is expected to be even better suited for heavy-deformed nuclei. Especially since it is known that the rotational law $J(J+1)$ is essentially generated by the angular momentum projection (AMP) of an intrinsic HF state (exemplified in Fig.~\ref{Cr48backbend}), which mainly minimizes the deformation component of the two-body interaction~\cite{Ripka1968}. A detailed comparison of this projected HF and the VAP calculations for nuclei near $^{254}$No has been recently presented in~\cite{No254ARXIV2025}, showing the overall good agreement between two calculations. We can thus conclude that:
\begin{itemize}
\item Strong pairing correlations can be adequately described using non-orthogonal Slater determinants when the spin-variation after projection is carried out ;
\item In well-deformed heavy nuclei where the deformation excels over the pairing effects, producing a nearly proper rotational sequence~\cite{Reiter1999_PhysRevLett.82.509,Greenlees2012_PhysRevLett.109.012501,Seweryniak2023_PhysRevC.107.L061302}, the use of Slater determinants within a VAP calculation can therefore be expected to be even more powerful than in light systems.  
\end{itemize}

\section{Converging the spherical ground state of $^{78}$Ni}
\label{sect5}
\begin{figure}[H]
  \centering
  \includegraphics[scale=1.6]{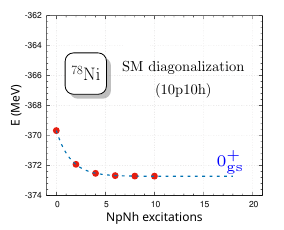}
  \includegraphics[scale=1.6]{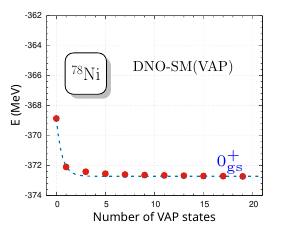}
  \caption{Binding energy ($E$ in MeV) convergence in $^{78}$Ni for the ground state $0^+_{\mathrm{gs}}$ with respect to $N$-particle $N$-hole level (SM diagonalization in top panel) and with the number of Slater determinants (DNO-SM(VAP)) in bottom panel). \label{Ni78}}
\end{figure}

The doubly magic $^{78}$Ni is one of the most exciting and challenging cases both experimentally and theoretically~\cite{Taniuchi2019_Ni78Nature}. From the shell-model perspective, it is representative of the largest feasible limit in large-scale shell-model diagonalizations using Lanczos algorithms~\cite{Caurier2005}.
This nucleus is described in the pf-sdg valence space using the PFSDG-U effective interaction built on top of $^{60}$Ca core~\cite{Taniuchi2019_Ni78Nature}. 
The full $M$-scheme dimension of Slater determinant basis is $210\:046\:691\:518 \sim 2\times 10^{11}$.  In Fig.~\ref{Ni78} (top panel), the binding energy convergence is shown for shell-model diagonalizations with respect to the $N$p$N$h excitations across Z=28 and N=50 shell gaps. At the $10$p$10$h level representing our actual diagonalization limit, the ground state energy is $-372.71668$ MeV. The corresponding extrapolated value ground state binding energy is, assuming an exponential convergence behavior~\cite{Caurier2005}, found to be $-372.72850$ MeV.
The energy convergence with the number of non-orthogonal VAP Slater states is shown in bottom panel of Fig.~\ref{Ni78} starting from a spherical Hartree-Fock solution (indicated at $0$ on the horizontal axis). An exponential-like behavior is observed as in the case of exact SM diagonalization. The converged energy in Fig.~\ref{Ni78} (bottom panel) is $-372.73275$ MeV, lower than the SM diagonalization value. This is the first time using a variational method, the exact diagonalization is fully recovered in such a challenging nucleus at the edge of current computational capacities. The prospective brought by this finding is very promising for the neutron-rich region centered around $^{78}$Ni where shape coexistence seems to be omnipresent~\cite{Nies2023_PhysRevLett.131.222503}. The control of the convergence will therefore allow a reliable description in this complex region.

\section{Conclusions and perspectives}
In summary, motivated by the formal proof of the Broeckhove-Deumens theorem on the completeness of non-orthogonal basis, the present work aims to find numerical evidences of this theorem in the resolution of realistic shell-model Hamiltonians. We solve this long-standing puzzle using symmetry-restored non-orthogonal Slater determinants as the variational ansatz under the Ritz variational principle. As an implication of the theorem, it is shown that non-orthogonal Slater determinants span the full shell-model valence space. In practice, their implementation under the angular-momentum variation after projection exhibits an outstanding performance to tackle the shell-model eigenvalue problem. The nature of the variation after projection is numerically demonstrated to be equivalent to promoting a huge number of many-particle many-hole excitations on top of deformed or spherical reference Slater states. In the case of strong pairing correlations as exemplified in $^{48}$Cr, the ground state correlations are fully recovered compared to the exact SM calculation. Excited states are similarly obtained with a dozen of basis states. These results show an important and non-trivial feature of non-orthogonal Slater determinants: in a strong pairing regime, proton-neutron pairing can be exactly incorporated without breaking the particle numbers. In the spherical regime, the description of the binding energy is also excellent as exemplified in the doubly-magic $^{78}$Ni. These findings open a new perspective to investigate heavy-mass nuclei as done in~\cite{No254ARXIV2025} where the use of Slater determinants can present a major computational simplification in the restoration of symmetries. Under such conditions, the VAP approach will also allow a critical assessment of the effective interaction, for example, in cases~\cite{Nies2023_PhysRevLett.131.222503} whose convergence is difficult in the conventional shell-model diagonalization. 


\section{Acknowledgements}
The authors acknowledge the financial support from CNRS/IN2P3,
France, ABI-CONFI Master project.

\appendix

\section{Broeckhove-Deumens theorem: symmetry-restored versus symmetry-breaking ansatz}
\label{APP}
We recall here the arguments in the proof of the Broeckhove-Deumens theorem~\cite{Deumens1979}. It is based on i) the separability of $\mathscr H = \overline{\mathrm{span} \: \mathit\Gamma}$ and ii) the continuous set $\mathit\Gamma$ is dense in $\mathscr H$. The first property i) means every element $\psi\in\mathscr H$ is written as an linear combination of an orthonormal basis $\xi_n$
\begin{equation}
\psi = \sum_{n=0}^{\infty} c_n\xi_n,
\end{equation}
so that there is some $K > 0$ satisfying
\begin{equation}
    \Big| \psi - \sum_{n=0}^K c_n\xi_n\Big| < \frac \epsilon 2
\end{equation}
for every $\epsilon$. The second property ii) implies the existence of a countable set $\mathit\Gamma_0 = \{\phi(\alpha_j(n,k))|j=0,...,M(n,k)\}\subset\mathit\Gamma$ such that for $\epsilon_k$ and $\xi_n$, one can find a linear combination which comes arbitrarily close to $\xi_n$, i.e.
\begin{equation}
\Big|\xi_n - \sum_{j=0}^{M(n,k)} a_j(n,k)\phi(\alpha_j(n,k))\Big| < \epsilon_k.
\end{equation}
A suitable choice of $\epsilon_k$ as done in~\cite{Deumens1979} and the use of the triangle inequality lead to 
\begin{equation}
    \Big|\psi - \sum_{n=0}^K \sum_{j=0}^{M(n,k)} c_n a_j(n,k)\phi(\alpha_j(n,k))\Big| < \epsilon.
\end{equation}
The set $\mathit\Gamma_0$ then forms a generally skew basis for $\mathscr H$. We now discuss the implications of the theorem in relation with symmetry-conserving and symmetry-breaking wave function ansatz. First, by construction the shell-model valence space $\mathscr H_{SM}$ is of finite dimensional. Secondly, its orthonormal basis ($\xi_n$ analogue) is spherical Slater determinants (the $M$-scheme basis~\cite{Caurier2005}). The Broeckhove-Deumens theorem in this context thus implies that
\begin{itemize}
\item There exists a countable set of non-orthogonal Slater determinants spanning the full shell-model space;
\item The number of non-orthogonal Slater determinants is finite.
\end{itemize}
These implications are guaranteed to hold regardless of whether the symmetries of the underlying Hamiltonian are restored or not. Now let us consider the triaxial coordinates $(\beta,\gamma)$~\cite{RingSchuck1980} often used to generate a space $\mathcal E_{def} = \overline{\mathrm{span}\:\it{\Gamma(\beta,\gamma)}}$ relevant for the ground-state deformation. The separability property is not obvious to be proved independently of physical cases, because of, for example, possible higher-order deformations which require an explicit inclusion of them. In general, the triaxial space $\mathcal E_{def}$ can thus be considered as a subspace of the shell-model space $\mathscr H_{SM}$. It is sufficient to verify this statement in a realistic example where the $(\beta,\gamma)$ degrees of freedom are physically the most favorable. This is the case for sd shell even-even nuclei (cf. Section~\ref{sect3} and Fig.~\ref{convergence}).  

In order to obtain a complete set of basis states, although the Broeckhove-Deumens theorem does not tell us how to find it, in this paper, we have shown that this task can be done with the Ritz variational principle applied to symmetry-restored Slater determinants. This variational calculation leads to the minimization of the projected energy~\eqref{Evap} through the energy gradient
\begin{equation}
   \begin{aligned} 
    G^{(q)}_{ij} = \frac{\displaystyle\sum_{K,q'K'} C^{*\pi J}_{n,qK}C^{\pi J}_{n,q'K'} \elmx{\phi_q}{a^\dagger_j a_i\big(\hat H - E^{\pi J}_n\big)\mathcal P^J_{KK'}P^\pi}{\phi_{q'}}}{\displaystyle\sum_{qK,q'K'}C^{*\pi J}_{n,qK}C^{\pi J}_{n,q'K'} \elmx{\phi_q}{\mathcal P^J_{KK'}P^\pi}{\phi_{q'}}}
  \end{aligned}
  \label{projected_Gij}
\end{equation}
In the un-projected ansatz, the gradient is of simpler form
\begin{equation}
   \begin{aligned} 
    g^{(q)}_{ij} = \frac{\displaystyle\sum_{q'} C^{*}_{n,q}C_{n,q'} \elmx{\phi_q}{a^\dagger_j a_i\big(\hat H - E_n\big)}{\phi_{q'}}}{\displaystyle\sum_{q,q'}C^{*}_{n,q}C_{n,q'} \overlap{\phi_q}{\phi_{q'}}}.
  \end{aligned}
  \label{unprojected_gij}
\end{equation}
The gradients~\eqref{projected_Gij},~\eqref{unprojected_gij} and Hamiltonian matrix elements  in~\eqref{secular_problem} can be evaluated by the Wick theorem~\cite{RingSchuck1980,UTSUNO2013}. The angular momentum projection is performed using the same method as~\cite{DNO2022_PhysRevC.105.054314}.  Without the restoration of symmetries, how efficient it is to solve the shell-model eigenvalue problem using~\eqref{unprojected_gij} remains an open practical problem. Finally, one notices that non-orthogonal Hartree-Fock-Bogoliubov wavefunctions should follow the same consequences stipulated by the Broeckhove-Deumens theorem, albeit with further complexity in practical calculations due to the particle number projection. We leave those questions for a future study.

\bibliographystyle{elsarticle-num}\biboptions{sort&compress}
\bibliography{main}






\end{document}